\documentstyle[amsfonts,preprint,aps]{revtex}
\begin{document}
\title{Teleportation scheme implementing the Universal Optimal Quantum Cloning
Machine and the Universal NOT gate}
\author{M. Ricci, F. Sciarrino, C. Sias and F. De Martini}
\address{Dipartimento di Fisica and \\
Istituto Nazionale per la Fisica della Materia\\
Universit\`{a} di Roma ''La Sapienza'', Roma, 00185 - Italy}
\maketitle

\begin{abstract}
The universal quantum cloning machine and the universal NOT\ gate acting on
a single qubit can be implemented very generally by slightly modifying the
protocol of quantum state teleportation. The experimental demonstration of
the $1\rightarrow 2$ cloning process according to the above scheme has been
realized for a qubit encoded in photon polarization.
\end{abstract}

Classical information is encoded in bits, viz. dichotomic variables that can
assume the values $0$ or $1.$ For such variables there are no theoretical
limitations as far as the ''cloning'' and/or ''spin-flipping'' processes are
concerned. However manipulations on the quantum analogue of a bit, a qubit,
have strong limitations due to fundamental requirements by quantum
mechanics. For instance, it has been shown that an arbitrary unknown qubit
cannot be perfectly cloned: $\left| \Psi \right\rangle \rightarrow \left|
\Psi \right\rangle \left| \Psi \right\rangle $, a consequence of the
so-called ``no cloning theorem'' \cite{1}, \cite{2}. Another ''impossible''
device is the quantum NOT gate, the transformation that maps any qubit into
the orthogonal one $\left| \Psi \right\rangle \rightarrow \left| \Psi
^{\perp }\right\rangle $ \cite{3}. In the last years a great deal of
theoretical investigation has been devoted to finding the best approximation
allowed by quantum mechanics to these two processes and to establish for
them the corresponding ''optimal'' values of the ''fidelity'' $F<1$.\ This
problem has been solved in the general case \cite{4,5,6}. In particular, it
was found that a one-to-two universal optimal quantum cloning machine
(UOQCM), i.e. able to clone one qubit\ into two qubits ($1\rightarrow 2$),
can be realized with a fidelity $F_{CLON}=\frac{5}{6}$. It has also been
proposed to test this result by a quantum optical ''amplification'' method,
i.e. by associating the cloning effect with the QED\ ''stimulated emission''
process \cite{7,8}. This proposal was quickly followed by some successful
experimental demonstrations \cite{9,10,11}. Very recently it has been argued
by \cite{11} that when the cloning process is realized in a subspace ${\it H}
$ of a larger nonseparable Hilbert space ${\it H}\otimes K$ which is acted
upon by a physical apparatus, the same apparatus performs contextually in
the space $K$ the ''flipping'' of the input injected qubit, then realizing a 
$(1\rightarrow 1)$ Universal-NOT gate (U-NOT) with a fidelity $F_{NOT}=\frac{%
2}{3}$ \cite{5,6}. As an example, a UOQCM can be realized on one output mode
of a non-degenerate ''quantum-injected'' optical parametric amplifier
(QIOPA), while the U-NOT transformation is realized on the other mode \cite
{12}. Precisely this QIOPA apparatus has been adopted very recently to
demonstrate experimentally the simultaneous, contextual realization of both
processes \cite{11,12}.

In the present work this relevant, somewhat intriguing result is
investigated under a new perspective implying a modified quantum state
teleportation (QST)\ protocol, according to the following scheme. The QST\
protocol implies that an unknown input qubit $\left| \phi \right\rangle
_{S}=\alpha \left| 0\right\rangle _{S}+\beta \left| 1\right\rangle _{S}\ $is
destroyed at a sending place (Alice:\ $A$) while its perfect replica appears
at a remote place (Bob:\ $B$) via dual quantum and classical channels. To
accomplish that two qualitatively different resources are needed: a quantum
resource that cannot be cloned, and a classical resource that cannot travel
faster than light \cite{13,14}. Let us assume that Alice and Bob share the
entangled ''singlet''\ state $\left| \Psi ^{-}\right\rangle _{AB}=2^{-%
%TCIMACRO{\UNICODE[m]{0xbd}}%
%BeginExpansion
{\frac12}%
%EndExpansion
}\left( \left| 0\right\rangle _{A}\left| 1\right\rangle _{B}-\left|
1\right\rangle _{A}\left| 0\right\rangle _{B}\right) $, and that we want to
teleport the generic qubit $\left| \phi \right\rangle _{S}$. The ''singlet''
state is adopted here and hereafter because its well known invariance under
SU(2) transformations will ensure the ''universality'' of the cloning and
U-NOT processes, as we shall see \cite{5,6,7,8,11,12}. The overall state of
the system is then $\left| \Omega \right\rangle _{SAB}=\left| \phi
\right\rangle _{S}\left| \Psi ^{-}\right\rangle _{AB}$. Alice performs a
Bell measurement by projecting the qubits $S$ and $A$ into the four Bell
states $\left\{ \left| \Psi ^{-}\right\rangle _{SA},\left| \Psi
^{+}\right\rangle _{SA},\left| \Phi ^{-}\right\rangle _{SA},\left| \Phi
^{+}\right\rangle _{SA}\right\} $ and then sends the result to Bob by means
of 2 bits of classical information. In order to obtain $\left| \phi
\right\rangle _{B}$, Bob applies to the received state the appropriate
unitary transformation $U_{B}$ according to the following protocol: $\left|
\Psi ^{-}\right\rangle _{SA}\rightarrow U_{B}={\Bbb I}$, $\left| \Psi
^{+}\right\rangle _{SA}\rightarrow U_{B}=\sigma _{Z}$, $\left| \Phi
^{-}\right\rangle _{SA}\rightarrow U_{B}=\sigma _{X}$, $\left| \Phi
^{+}\right\rangle _{SA}\rightarrow U_{B}=\sigma _{Y}$ where the kets express
the received corresponding information,$\ I,\sigma _{Z},\sigma _{X}$ are
respectively the identity, phase flip, spin flip operators and $\sigma
_{Y}=-i\sigma _{Z}\sigma _{X}$. The teleportation channel, after the unitary
operation $U_{B}$, acts as the identity operator: $E_{QST}(\rho _{S})={\Bbb %
\rho }_{B}$ where $\rho _{S}$ is the density matrix representing the state
of the qubit $S$. In absence of the classical information channel, Bob
cannot apply the appropriate $U_{B}$, and the apparatus realized the map $%
E_{B}(\rho _{S})=\frac{{\Bbb I}_{B}}{2}$, corresponding to the depolarizing
channel $E_{DEP}$. This is the worst possible information transfer because
any information about the initial state $\left| \phi \right\rangle $ is lost.

In order to implement the UOQCM and UNOT machines at Alice's and Bob's sites
respectively, in the present letter we modify the QST\ protocol by
performing a different projective measurement on the systems $S$ and $A$.
This leads to a different classical information to be transferred from $A$
to $B$: Fig.1. Precisely, the complete measurement able to discriminate
between the four Bell states is replaced by a dichotomic measurement able to
identify $\left| \Psi ^{-}\right\rangle _{SA}$ and its complementary space.
Let us analyze the outcomes of such strategy. With a probability $p=\frac{1}{%
4}$ the state $\left| \Psi ^{-}\right\rangle _{SA}$ is detected by $A$. In
this case the correct QST channel $E_{QST}$ is realized. If\ with
probability $p=\frac{3}{4}\ $this is not the case, Bob does not apply any
unitary transformation to the set of non identified Bell states $\left\{
\left| \Psi ^{+}\right\rangle _{SA},\left| \Phi ^{-}\right\rangle
_{SA},\left| \Phi ^{+}\right\rangle _{SA}\right\} $, and then the QST
channel implements the statistical map $E\left( \rho \right) =\frac{1}{3}%
\left[ \sigma _{Z}\rho \sigma _{Z}+\sigma _{X}\rho \sigma _{X}+\sigma
_{Y}\rho \sigma _{Y}\right] $. This map indeed coincides with the map $%
E_{UNOT}\left( \rho _{S}\right) $ which realizes the Universal Optimal NOT
gate, i.e. the one that approximates ''optimally'' the flipping of one qubit 
$\left| \phi \right\rangle $ into the orthogonal qubit $\left| \phi ^{\perp
}\right\rangle $. Bob identifies the two different maps realized at his site
by reading the information (1 bit) received by Alice on the classical
channel. For example, such bit can assume the value $0$ if Alice identifies
the Bell state $\left| \Psi ^{-}\right\rangle _{SA}$ and $1$ if not.

As already stated, in a bipartite entangled system the U-NOT gate is
generally realized with an optimal quantum cloning process \cite{11,12}.
Therefore it is worth analyzing carefully what happens when the overall
state $\left| \Omega \right\rangle _{SAB}$\ is projected onto the subspace
orthogonal to $\left| \Psi ^{-}\right\rangle _{SA}\left\langle \Psi
^{-}\right| _{SA}\otimes H_{B}$, i.e. by the projector: 
\begin{equation}
P_{SAB}=({\Bbb I}_{SA}-\left| \Psi ^{-}\right\rangle _{SA}\left\langle \Psi
^{-}\right| _{SA})\otimes {\Bbb I}_{B}  \label{proiettore}
\end{equation}
\newline
It generates the normalized projected state $\left| \widetilde{\Omega }%
\right\rangle \equiv P_{SAB}\left| \Omega \right\rangle _{SAB}=\sqrt{\frac{2%
}{3}}[\left| \xi _{1}\right\rangle _{SA}\otimes \left| 1\right\rangle
_{B}-\left| \xi _{0}\right\rangle _{SA}\otimes \left| 0\right\rangle _{B}]$
where $\left| \xi _{1}\right\rangle _{SA}=\alpha \left| 0\right\rangle
_{S}\left| 0\right\rangle _{A}+\frac{\beta }{2}\left| 1\right\rangle
_{S}\left| 0\right\rangle _{A}+\frac{\beta }{2}\left| 0\right\rangle
_{S}\left| 1\right\rangle _{A}$ and $\left| \xi _{0}\right\rangle
_{SA}=\beta \left| 1\right\rangle _{S}\left| 1\right\rangle _{A}+\frac{%
\alpha }{2}\left| 1\right\rangle _{S}\left| 0\right\rangle _{A}+\frac{\alpha 
}{2}\left| 0\right\rangle _{S}\left| 1\right\rangle _{A}$. By projecting
this state on the manifold $SA$ the following density matrix is obtained: 
\begin{equation}
\rho _{SA}\equiv Tr_{B}\left| \widetilde{\Omega }\right\rangle \left\langle 
\widetilde{\Omega }\right| =\frac{2}{3}\left| \phi \right\rangle \left| \phi
\right\rangle _{SA}\left\langle \phi \right| \left\langle \phi \right| _{SA}+%
\frac{1}{3}\left| \left\{ \phi ,\phi ^{\perp }\right\} \right\rangle
_{SA}\left\langle \left\{ \phi ,\phi ^{\perp }\right\} \right| _{SA}
\label{RhoSA}
\end{equation}
where $\left| \left\{ \phi ,\phi ^{\perp }\right\} \right\rangle _{SA}=2^{-%
%TCIMACRO{\UNICODE[m]{0xbd}}%
%BeginExpansion
{\frac12}%
%EndExpansion
}\left( \left| \phi ^{\perp }\right\rangle _{S}\left| \phi \right\rangle
_{A}+\left| \phi \right\rangle _{S}\left| \phi ^{\perp }\right\rangle
_{A}\right) $. We further project on the reduced spaces obtaining the
matrices:\newline
\begin{equation}
\rho _{S}\equiv Tr_{A}\rho _{SA}=\frac{5}{6}\left| \phi \right\rangle
\left\langle \phi \right| +\frac{1}{6}\left| \phi ^{\perp }\right\rangle
\left\langle \phi ^{\perp }\right| =\rho _{A}\equiv Tr_{S}\rho _{SA}
\end{equation}

\begin{equation}
\rho _{B}\equiv Tr_{SA}\left| \widetilde{\Omega }\right\rangle \left\langle 
\widetilde{\Omega }\right| =\frac{2}{3}\left| \phi ^{\perp }\right\rangle
\left\langle \phi ^{\perp }\right| +\frac{1}{3}\left| \phi \right\rangle
\left\langle \phi \right|
\end{equation}
by which the ''optimal'' values for the fidelities of the two ''forbidden''\
processes are obtained: $F_{CLON}=\frac{5}{6}$ and $F_{UNOT}=\frac{2}{3}\ $%
\cite{11,12}.

We want also to emphasize that it is possible to obtain the same results by
performing a projective measurement on different Bell states. It's easily
verified that by projection along $({\Bbb I}_{SA}-\left| \Phi
^{+}\right\rangle _{SA}\left\langle \Phi ^{+}\right| _{SA})\otimes {\Bbb I}%
_{B}$, the teleportation channel realizes the optimal universal
approximation of the transpose map. This map is related to the UNOT map by a
unitary transformation: $E_{TRANSPOSE}=E_{\sigma _{Y}}\left( E_{UNOT}\left(
\rho \right) \right) $, where $E_{\sigma _{Y}}\left( \rho \right) =\sigma
_{Y}\rho \sigma _{Y}$. It's straightforward verified that the ''cloning
machine'', in this case, generates $\rho _{S}^{\prime }=\rho _{A}^{\prime
}=\sigma _{Y}\rho _{S}\sigma _{Y}$, that is, the clones of the input qubit
on which acts the unitary transformation $\sigma _{Y}$. Similarly, by
choosing another Bell state in (\ref{proiettore}), analogue transformations
are achieved. Of course, in this case the universality of the processes is
retrieved by applying an adequate unitary transformation to the output
states of qubits labelled by $SA$ and $B$.

Note that in the present context the presence of the entangled state $\left|
\Psi ^{-}\right\rangle _{AB}\ $is not strictly necessary for the
implementation of the solely quantum cloning process. Indeed for this
purpose we could inject into the Alice apparatus in Fig.2 the input qubit $%
\left| \phi \right\rangle $ together with a {\it fully mixed }state $\rho
_{A}=\frac{{\Bbb I}_{A}}{2}$ spanning a $2$ dimension space. Of course in
this case only the Alice's apparatus is relevant and the U-Not process is
absent. This is the solution chosen for our experimental demonstration:\
Fig.2. The qubit to be cloned is: $\left| \phi \right\rangle _{S}$ $=\alpha
\left| H\right\rangle _{S}+\beta \left| V\right\rangle _{S}$ where $\left|
H\right\rangle $ and $\left| V\right\rangle $ respectively correspond to the
horizontal and vertical linear polarizations of a single photon injected in
one input mode $S$ of a $50:50\,$beamsplitter $BS_{A}$. A fully mixed state $%
\rho _{A}$ is simultaneously injected on the other input mode $A$ of $BS_{A}$
where the two input modes are lineraly superimposed. Consider the overall
output state which is realized on the two ouput modes $1\,$and $2$ of $%
BS_{A} $. It can be expressed as a linear combination of the Bell states $%
\left\{ \left| \Psi ^{-}\right\rangle _{SA},\left| \Psi ^{+}\right\rangle
_{SA},\left| \Phi ^{-}\right\rangle _{SA},\left| \Phi ^{+}\right\rangle
_{SA}\right\} $. As it is well known, the realization of the singlet $\left|
\Psi _{SA}^{-}\right\rangle $ is unambiguously indentified by the detection
of two single photons on the output modes of $BS_{A}$ while the realization
of the set of the other three Bell states correspond to the emission of
photon pairs on either one of the output modes \cite{15}. Hence the
detection of two photons over either the mode $1$ or $2$, a
Bose-mode-occupation enhancement, implies the projection by $P_{SA}=\left( 
{\Bbb I}_{SA}-\left| \Psi _{SA}^{-}\right\rangle \left\langle \Psi
_{SA}^{-}\right| \right) $ of the output state into the space orthogonal to $%
\left| \Psi _{SA}^{-}\right\rangle $ \cite{16}. As seen in the theoretical
analysis above this condition leads, in the general case, to the
simultaneous realization of UOQCM and UNOT by a post-selection technique.

In the present experiment, a pair of non entangled photons with wavelength $%
\lambda =532nm$ and with a coherence-time $\tau _{coh}=80fs$, were generated
by a spontaneous parametric down conversion SPDC process in a Type I BBO\
crystal in the initial polarization product state $\left| H\right\rangle
_{S}\left| H\right\rangle _{A}$. The two photons were then injected on the
two input modes $S$ and $A$ of $BS_{A}$. The input qubit $\left| \phi
\right\rangle _{S}$, associated with mode $S$ was polarization encoded by
means of a waveplate (wp) $WP_{S}$. The transformation used to map the state 
$\left| H\right\rangle _{A}$ into $\rho _{A}=\frac{{\Bbb I}_{A}}{2}$ was
achieved by stochastically rotating a $\lambda /2$ waveplate inserted on the
mode $A$ during the experiment. In this way the statistical evolution of $%
\left| H\right\rangle _{A}$ into two orthogonal states with equal
probability was achieved. The photons $S$ and $A$ were injected in the two
input arms of $BS_{A}$ with a mutual delay $\Delta t$ micrometrically
adjustable by a translation stage with position settings $Z=2\Delta tc$. The
setting value $Z=0$ was assumed to correspond to the full overlapping of the
photon pulses injected into $BS_{A}$, i.e. to the maximum photon
interference.

For the sake of simplicity, we only analyzed the measurements performed on
the $BS_{A}$ output mode $1$:\ Fig.2. The polarization state on this mode
was analyzed by the combination of the wp $WP_{C}$ and of the polarization
beam splitter $PBS_{C}$. For each input polarization state $\left| \phi
\right\rangle _{S}$, $WP_{C}$ was set in order to make $PBS_{C}$ to transmit 
$\left| \phi \right\rangle _{S}$ and reflect $\left| \phi ^{\perp
}\right\rangle _{S}$. The ''cloned'' state $\left| \phi \phi \right\rangle
_{S}$ could be detected on mode $1$ by a two-photon counter, realized in our
case by first separating the two photons by a $50:50$ beam splitter $BS_{C}$
and then detecting the coincidence $[D_{A1},D_{A2}]$ between the output
detectors $D_{A1}$ and $D_{A2}$:\ Fig.2. Any coincidence between detectors $%
D_{B}$ and $D_{A2}$ corresponded to the realization of the state $\left|
\phi \phi ^{\perp }\right\rangle _{S}$. First consider the cloning machine
switched off, by setting: $\Delta t>\tau _{coh\text{\ }}$i.e. by making $S$
and $A$ not interfering on $BS_{A}$. In this case, since the states $\left|
\phi \phi \right\rangle _{S}$ and $\left| \phi \phi ^{\perp }\right\rangle
_{S}$ were realized with the same probability on mode $1$, the rate of
coincidences detected by $[D_{A1},D_{A2}]$ and $[D_{A2},D_{B}]\ $were
expected to be equal. By turning on the cloning machine, i.e. by setting $%
\Delta t<<\tau _{coh}$, on mode $1$ the output density matrix $\rho _{SA}$
was realized implying an enhancement by a factor $R=2$ of the counting rate
by $[D_{A1},D_{A2}]$ and no rate enhancement by $[D_{A2},D_{B}]$. The
measurement of $R$ was carried out by coincidence measurements involving
simultaneously $[D_{A1},D_{A2}]$ and $[D_{A2},D_{B}]$. The experimental data
are reported\ in Figure 3 for three different input states $\left| \phi
\right\rangle _{S}=\left| H\right\rangle $, $2^{-%
%TCIMACRO{\UNICODE[m]{0xbd}}%
%BeginExpansion
{\frac12}%
%EndExpansion
}(\left| H\right\rangle +\left| V\right\rangle )$, $2^{-%
%TCIMACRO{\UNICODE[m]{0xbd}}%
%BeginExpansion
{\frac12}%
%EndExpansion
}(\left| H\right\rangle +i\left| V\right\rangle )$. There circles and square
marks refer respectively to the $[D_{A1},D_{A2}]$ and $[D_{A2},D_{B}]$
coincidences versus the position setting $Z.$ We may check that the cloning
process only affects the $\left| \phi \phi \right\rangle _{S}$ component, as
expected and $R\;$is determined as the ratio between the peak values
(cloning machine switched on) and the basis values (cloning machine off).
The corresponding experimental values of the {\it cloning fidelity }$F=\frac{%
2R+1}{2R+2}$ are: $F_{H}=0.827\pm 0.002$;$\ F_{H+V}=0.825\pm 0.002$; $%
F_{H+iV}=0.826\pm 0.002$. These ones are to be compared with the optimal
value $F_{th}=5/6\approx 0.833$ which corresponds to the limit value of the
amplification ratio $R=2$. As a lucky aspect of the present work, note the
relevance in the theory above of the ''singlet'' state $\left| \Psi
_{SA}^{-}\right\rangle $ in our measurement procedure. As already remarked,
according to quantum theory the exclusive SU(2) invariance of this state
should imply the insensitivity of the value of the ''fidelity'' $F\ $to the
photon polarization, i.e. to the input state $\left| \phi \right\rangle _{S}$%
. Indeed the experimental results just reported expressing very close values
of $F$ for different polarization states confirm the ''{\it universality}''
of the cloning process.

In summary, two relevant quantum information processes, forbidden by quantum
mechanics in their exact form are found to be connected by a modified
quantum state teleportation scheme and can be ''optimally'' realized by a
simple linear method. As a partial implementation of the new protocol, an
experimental demonstration of one to two universal quantum cloning has been
performed by a quantum optical method. The results are found in good
agreement with the theoretical expectations. We believe that the new
perspective established by the present work will substantially enlighten an
important and still unclear aspect of fundamental quantum theory.

This work has been supported by the FET European Network on Quantum
Information and Communication (Contract IST-2000-29681: ATESIT), by Istituto
Nazionale per la Fisica della Materia (PRA\ ''CLON'')\ and by Ministero
dell'Istruzione, dell'Universit\`{a} e della Ricerca (COFIN 2002).

Figure.1. General scheme for the simultaneous realization of the Universal
Quantum Cloning Machine (UOQCM) and of the UNOT gate.

Figure.2. Experimental setup for the optical implementation of the UOQCM by
a modified Teleportation protocol (left part of the Figure).

Figure.3.\ Experimental result of the universal cloning process for
different input qubits corresponding to the encoded polarizations: $\left|
H\right\rangle $, $2^{-1/2}\left( \left| H\right\rangle +\left|
V\right\rangle \right) $ and $2^{-1/2}\left( \left| H\right\rangle +i\left|
V\right\rangle \right) $.

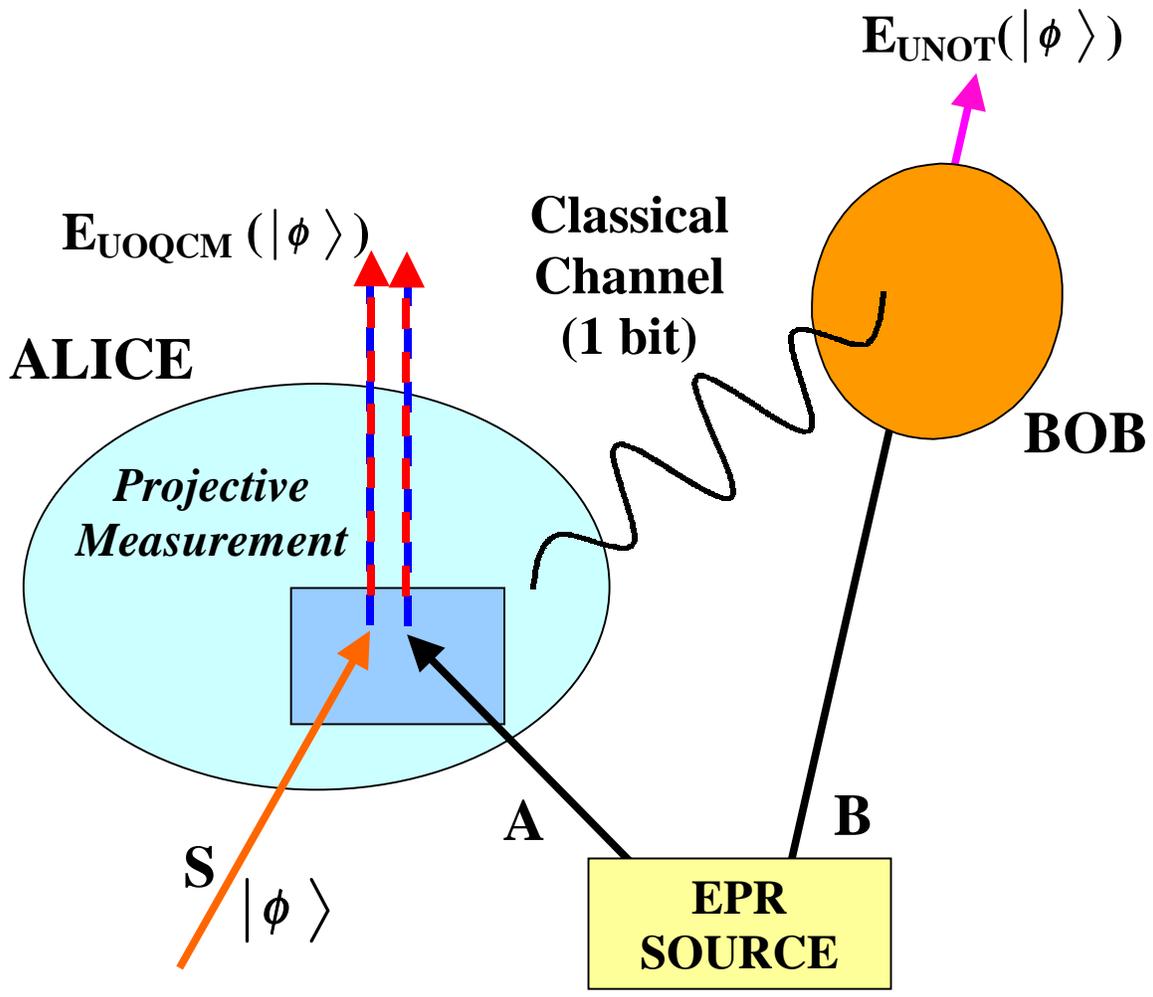
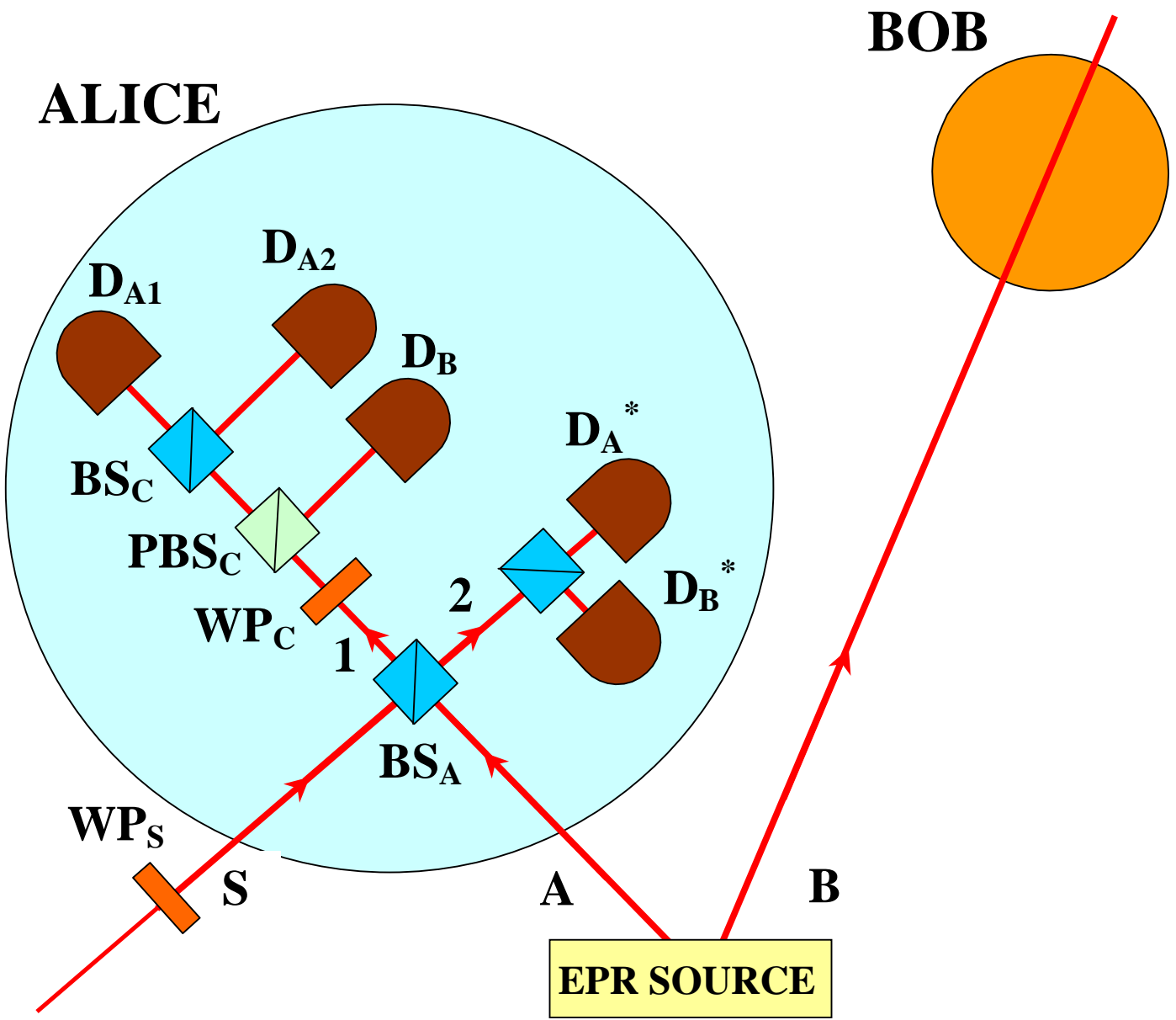
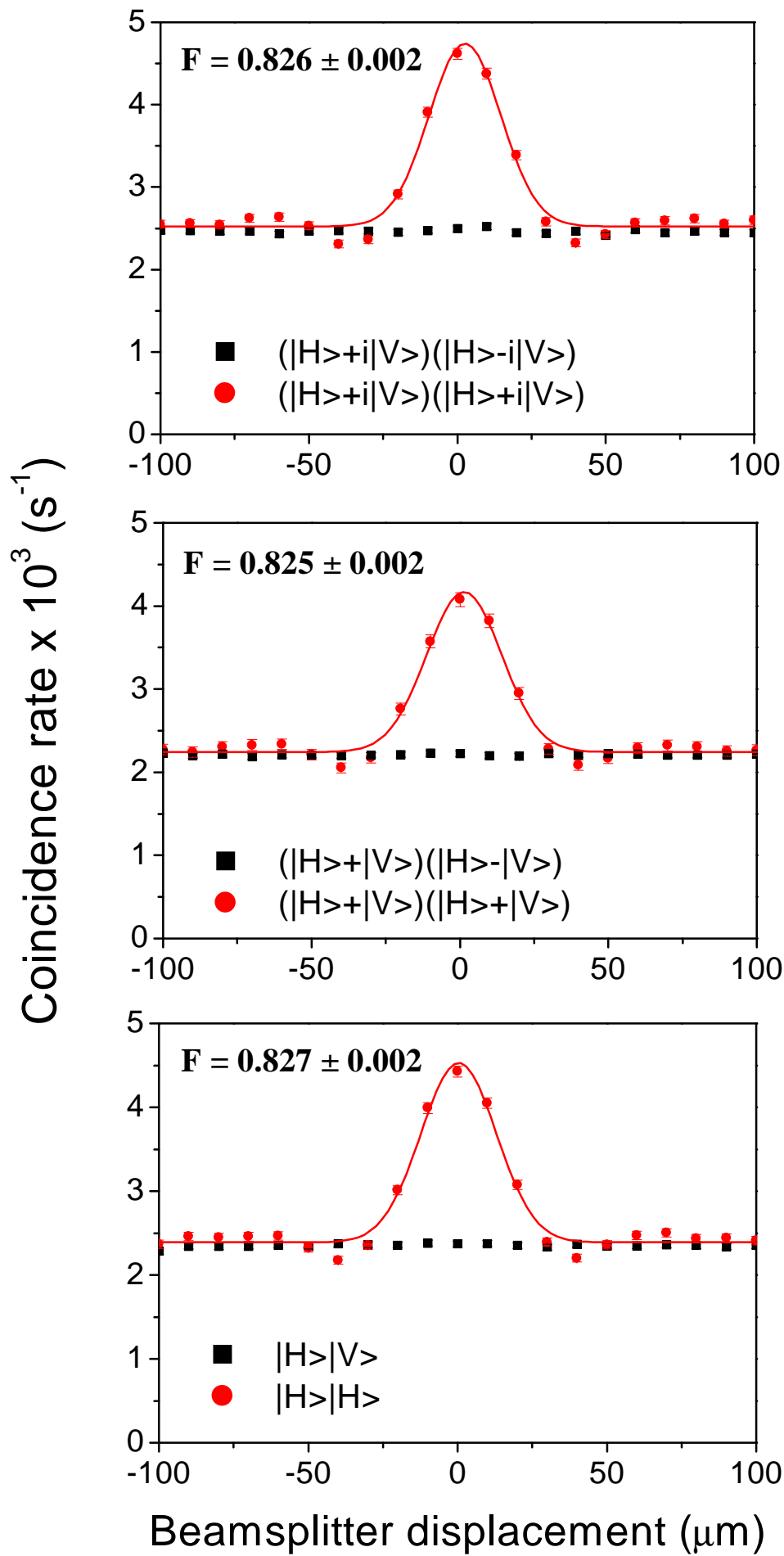
\end{document}